\documentclass[prd,aps,twocolumn,a4paper,floatfix,nofootinbib]{revtex4-1}   

\usepackage[utf8]{inputenc}
\usepackage{graphicx,psfrag}
\usepackage{mathrsfs}
\usepackage{amsmath,amsfonts,amssymb}
\usepackage{multirow} 
\usepackage{diagbox}
\usepackage{comment}
\usepackage{xcolor}
\usepackage{enumerate}
\usepackage{booktabs}

\usepackage{hyperref}
\hypersetup{
    colorlinks = true,
    linkcolor = {blue},
    citecolor = {blue},
    urlcolor = {blue},
    linkbordercolor = {white},
    }

\usepackage{color}
\definecolor{cyan}{rgb}{0,0.9,0.9}
\definecolor{orange}{rgb}{0.9,0.5,0}
\definecolor{magenta}{rgb}{1,0,1}
\definecolor{purple}{rgb}{0.8,0.4,0.8}
\definecolor{gray}{rgb}{0.8242,0.8242,0.8242}
\definecolor{mgreen}{rgb}{0.1,0.8,0.1}

\usepackage[normalem]{ulem}

\begin{document}

\title{Impact of gravitational waveform model systematics on the measurement of the Hubble constant}

\author{Nina \surname{Kunert}$^{1}$}
\email{nkunert@uni-potsdam.de}
\author{Jonathan~\surname{Gair}$^{2}$}
\author{Peter~T.~H. \surname{Pang}$^{3,4}$}
\author{Tim \surname{Dietrich}$^{1,2}$}

\affiliation{${}^1$Institute for Physics and Astronomy, University of Potsdam, D-14476 Potsdam, Germany}
\affiliation{${}^2$Max Planck Institute for Gravitational Physics (Albert Einstein Institute), Am M\"uhlenberg 1, Potsdam 14476, Germany}
\affiliation{${}^3$Institute for Gravitational and Subatomic Physics (GRASP), Utrecht University, Princetonplein 1, 3584 CC Utrecht, The Netherlands}
\affiliation{${}^4$Nikhef, Science Park 105, 1098 XG Amsterdam, The Netherlands}

\date{\today}

\begin{abstract}

Matching gravitational-wave observations of binary neutron stars with theoretical model predictions reveals important information about the sources, such as the masses and the distance to the stars. The latter can be used to determine the Hubble constant, the rate at which the Universe expands. One general problem of all astrophysical measurements is that theoretical models only approximate the real underlying physics, which can lead to systematic uncertainties introducing biases. However, the extent of this bias for the distance measurement due to uncertainties of gravitational waveform models is unknown. In this study, we analyze a synthetic population of 38 binary neutron star sources measured with Advanced LIGO and Advanced Virgo at design sensitivity. We employ a set of four different waveform models and estimate model-dependent systematic biases on the extraction of the Hubble constant using the bright siren method. Our results indicate that systematic biases are below statistical uncertainties for the current generation of gravitational-wave detectors. 

\end{abstract}

\maketitle

\section{Introduction}
\label{sec:intro}

The measurement of the local expansion rate of the Universe, also known as the Hubble constant, $H_0$, is one of the most intriguing problems in cosmology. The knowledge of the cosmological expansion rate at different times, given through the Hubble-Lema\^{i}tre parameter, is essential for understanding the history of the Universe and to infer its age and composition. 

A big puzzle in cosmology is to piece together the measurements from the early and late Universe. While the SHOES Collaboration~\cite{Riess:2019cxk} using the cosmic distance ladder method constrained the local expansion rate to $H_0 = 74.03 \pm 1.42 \rm \ km \ s^{-1}Mpc^{-1}$, the Planck Collaboration~\cite{Planck:2018vyg} estimated $H_0 = 67.4 \pm 0.5 \rm \ km \ s^{-1}Mpc^{-1}$ at 68\% credibility from observations of the cosmic microwave background (CMB) radiation. This leads to the well-known Hubble-Lema\^{i}tre tension with a discrepancy at a $4.4$ sigma level~\cite{Riess:2019cxk}. 

This discrepancy has spurred scientific efforts, with researchers proposing numerous independent studies to elucidate the underlying cause. The use of acoustic baryonic oscillations~\cite{DES:2018rjw}, the gravitational lensing method~\cite{Birrer:2018vtm}, or calibrating the tip of the red giant branch (TRGB) to Type Ia supernovae~\cite{Freedman:2019jwv} exemplify prominent attempts to resolve this enigma. Meanwhile, the possibility of a need for new physics is actively debated, suggesting that a new cosmological model may be necessary to reconcile the distinct measurements~\cite{Berbig:2020wve, Blinov:2021mdk, Mandal:2023bzo, Poulin:2023lkg}. 

Using gravitational-wave (GW) observations as a probe to infer the cosmic expansion rate was first proposed by Schutz in 1986~\cite{Schutz1986}. The GW emission of compact binary coalescences involving black holes or neutron stars can serve as cosmic distance indicators or so-called standard sirens and provide a measurement of the luminosity distance of the source. The landmark detection of GW170817 \cite{LIGOScientific:2017vwq, LIGOScientific:2018hze}, a binary neutron star (BNS) merger, by the LIGO and Virgo detectors~\cite{LIGOScientific:2014pky, VIRGO:2014yos} along with its associated electromagnetic (EM) counterparts, GRB170817A and AT2017gfo~(\cite{LIGOScientific:2017ync} and refs.~therein), marked a pivotal milestone in the field of cosmology. This event enabled the first direct measurement of the Hubble constant using a GW standard siren~\cite{LIGOScientific:2017adf, LIGOScientific:2019zcs}. Although current constraints from GW observations are not yet able to resolve the Hubble tension, future observations could place further constraints on this cosmological parameter~\cite{Chen:2017rfc, Feeney:2018mkj, Farr:2019twy, Feeney:2020kxk}. Multi-messenger observations especially hold significant promise for improving the accuracy of Hubble constant determinations. This advantage arises from being able to combine the GW measurement of the distance to the source with an electromagnetic measurement of its redshift~\cite{LIGOScientific:2017adf, Guidorzi:2017ogy, Hotokezaka:2018dfi, Wang:2020vgr, Dietrich:2020efo, Bulla:2022ppy, Palmese:2023beh}.

While the standard siren methodology offers the advantage of not requiring recalibration, as is common practice in the distance ladder method, there are different sources of systematic uncertainties that must be addressed. Previous studies such as Ref.~\cite{huang2022impact} investigated the impact of instrumental calibration uncertainties on Hubble constant measurements. They found that for single BNS events with a signal-to-noise ratio (SNR) of 50 for a network consisting of Advanced LIGO and Advanced Virgo operating at design sensitivity, the introduced bias is smaller than statistical uncertainties, but could accumulate as more events are detected. Reference~\cite{Turski:2023lxq} examined the influence of galaxy redshift uncertainties using different galaxy redshift uncertainty models, with particular emphasis on photometric redshifts, which are known to exhibit substantial errors. Their findings indicate that while there may exist a potential bias, it remains smaller than statistical uncertainties. Furthermore, systematic uncertainties arising from the viewing angle of BNS mergers and EM selection effects can be major challenges \cite{Chen:2020dyt, Chen:2023dgw, mancarella2024accurate}. Building on \cite{2024arXiv240405811D}, this work explores the systematic biases introduced by binary black hole (BBH) waveform models. The analysis reveals that these biases become more pronounced with increasing detector-frame total mass, binary asymmetry, and spin-precession effects. Notably, the study demonstrates on the basis of three high SNR events (golden events) that current waveform models lead to biased measurement of the Hubble constant, even for current detectors. 

In contrast to Ref.~\cite{2024arXiv240405811D}, our study delves into the issue of systematic uncertainties originating from BNS waveform models potentially influencing the determination of the distance to the source and, hence, the Hubble constant. Until now, this question remains, up to the best of our knowledge, unanswered. We analyze a synthetic population of 38 BNS sources, previously published in Ref.~\cite{Kunert:2021hgm}, with a set of four different waveform models and study model-dependent systematic errors. We employ the most recent Planck CMB measurement \cite{Planck:2018vyg} as a benchmark and investigate whether estimating $H_0$ using different models than the one used to generate the data introduces a systematic bias. As our results will demonstrate, GW model-dependent biases have a minimal influence on the inferred value of the Hubble constant for current detector networks.

This paper is organized as follows. Section~\ref{sec:methods} provides details on modeling GW standard sirens, the simulated BNS source population, and on the Bayesian framework employed to extract estimates of $H_0$. Our results of combined $H_0$ estimates are presented in Sec.~\ref{sec:results} and Sec.~\ref{sec:conclusion} summarizes the key takeaways.

\section{Methodology}\label{sec:methods}

The standard siren method utilizes the detection of GW signals and their correlation with EM observations of the same astrophysical events to infer the Hubble constant. When a source's redshift can be measured directly or through an association with its host galaxy, it is denoted as a bright siren. When an EM counterpart is absent, statistical methods can be used to ascertain possible redshifts of the source. Notable examples of these techniques include the galaxy catalog method~\cite{PhysRevD.86.043011, Chen:2017rfc,LIGOScientific:2018gmd, Gray:2019ksv,LIGOScientific:2021aug, Gray2022,Gray:2023wgj,Mastrogiovanni:2023emh, Borghi:2023opd}, the cross correlation method~\cite{PhysRevD.93.083511, Mukherjee:2020hyn, Bera:2020jhx}, and the spectral siren technique~\cite{Farr_2019, Mastrogiovanni:2021wsd, Mukherjee:2021rtw, Ezquiaga_2022}. 

In our study, we will employ the bright siren technique to infer a joint estimate of the local expansion rate of the Universe, assuming the standard Lambda cold dark matter ($\Lambda$CDM) cosmology. In accordance, we utilize the latest Planck CMB measurement \cite{Planck:2018vyg} as a reference value for our Bayesian analysis.

\subsection{Modeling standard sirens}\label{subsec:GW_SSs}

Gravitational waveform models describe the amplitude $\Tilde{A}$ and the phase $\phi$ of standard sirens emitted from coalescing compact objects like neutron stars and black holes. The evolution of a gravitational waveform $\tilde{h}(f)$ over frequency can be written as
\begin{equation}\label{GW_freq}
    \tilde{h}(f) = \Tilde{A}(f) \exp(-i \phi(f)), 
\end{equation}
\noindent
in which both, the amplitude as well as the phase carry information about the intrinsic source properties. GW detectors measure a linear combination of two polarizations of the GW signal. At leading order, these can be written as \cite{Gray:2019ksv}
\begin{equation}\label{Eq:hplus}
    \tilde{h}_{+}(f) \propto \frac{\mathcal{M}_z^{5/6}}{D_L}\frac{1 + \cos^2(\iota)}{2} f^{-7/6} \exp({i \phi(\mathcal{M}_z,f)}),
\end{equation}
\begin{equation}\label{Eq:hcross}
    \tilde{h}_{\times}(f) \propto \frac{\mathcal{M}_z^{5/6}}{D_L} \cos{\iota} f^{-7/6} \exp({i \phi(\mathcal{M}_z, f) + i\pi/2}),
\end{equation}
\noindent
where $\mathcal{M}_z \equiv \mathcal{M}(1+z)$ denotes the detector-frame (or ``red-shifted'') chirp mass, which is better constrained by the signal's phase as compared its amplitude. It can be seen that the amplitude of a GW signal decreases in proportion to the luminosity distance $D_L$ of its source. This inverse relationship is a fundamental property of GWs and plays a crucial role in their detection and interpretation. On the other hand, gravitational waveform models including only the quadrupole mode show a limited capability to accurately infer the luminosity distance to an astrophysical source due to its interdependence with the inclination angle $\iota$. This degeneracy fundamentally limits our ability to measure both parameters so that the dominant uncertainty in the signal's amplitude arises from the uncertainties on the luminosity distance and the inclination angle. The inclusion of higher-order modes can be a way to break this degeneracy, as shown in Ref.~\cite{CalderonBustillo:2020kcg}.

\subsection{Population of synthetic BNS sources}\label{subsec:injsetup}

In this study, we use the large set of mock data obtained in Ref.~\cite{Kunert:2021hgm} and exploit the inferred posterior on the luminosity distance to study the impact of model uncertainties on the estimation of the Hubble constant. Below, we recap the key elements of the study.

The population of $38$ synthetic binary neutron star sources was simulated with stationary Gaussian noise into a network of interferometers assuming Advanced LIGO \cite{TheLIGOScientific:2014jea} and Advanced Virgo \cite{TheVirgo:2014hva} design sensitivity. In light of current observations suggesting a uniform distribution for neutron star masses in gravitational-wave binaries \cite{Landry:2021hvl}, we adopt a uniform prior sampled within $m_{1,2} \in [1,2]~M_{\odot}$. All sources were uniformly sampled in comoving volume with an optimal network SNR, $\rho$, ranging within $\rho \in [7, 100]$ and lie within a maximum or threshold luminosity distance of $D^{\rm thr}_L = 100$~Mpc. Hence, the population lies within a volume in which cosmology can be well modeled under the linear Hubble relation (for redshifts $z \ll 1$) given as
\begin{equation}\label{Eq:linH0relation}
    D_L (z) \approx \frac{cz}{H_0},
\end{equation}
\noindent
in which $c$ is the speed of light. For all BNS signals in our population, we employed \textsc{IMRPhenomD\_NRTidalv2} (PhenDNRTv2)~\cite{Dietrich:2019kaq} as our reference model that we use for the creation of the simulated data. We performed parameter estimation runs using \textsc{parallel bilby}~\cite{Smith:2019ucc} for all sources in the population with four different models, which were \textsc{TaylorF2} (TF2)~\cite{Sathyaprakash:1991mt,Blanchet:1995ez,Damour:2001bu,Blanchet:2004ek,Blanchet:2005tk,Mishra:2016whh, Mikoczi:2005dn, Arun:2008kb, Bohe:2015ana, Mishra:2016whh, Damour:2009wj,Vines:2011ud,Bini:2012gu,Damour:2012yf}, \textsc{IMRPhenomD\_NRTidal} (PhenDNRT)~\cite{Husa:2015iqa, Khan:2015jqa, Taracchini:2013rva, Dietrich:2017aum, Dietrich:2018uni, Nagar:2018zoe, Bernuzzi:2014owa}, \textsc{IMRPhenomD\_NRTidalv2} (PhenDNRTv2), and \textsc{SEOBNRv4\_ROM\_NRTidalv2} (SEOBNRTv2)~\cite{Bohe:2016gbl, Purrer:2014fza, Dietrich:2019kaq} resulting in a total of 152 simulations. For further details on our simulation setup, we refer the interested reader to Ref.~\cite{Kunert:2021hgm}.

\subsection{Bayesian framework}\label{subsec:Bayframework}

In this section, we outline our method for estimating the Hubble constant by utilizing the posterior distributions on luminosity distance obtained for the discussed BNS population. In order to compute the posterior probability distribution of the Hubble constant for a set of GW events $\left\{x_{\rm{GW}} \right\}$, we employ Bayes' theorem
\begin{equation}\label{Eq:Bayestheorem}
    p(H_0|\left\{x_{\rm{GW}}\right\}) \propto  p(H_0) p(\left\{x_{\rm{GW}} \right\}|H_0)
\end{equation}
\noindent
where $p(H_0)$ is the prior on $H_0$ and $p(\left\{x_{\rm{GW}}\right\}|H_0)$ is the likelihood of the measured GW data given a certain value of $H_0$.

Our analysis follows the bright siren approach, which utilizes the joint information from GW detections $\left\{x_{\rm{GW}} \right\}$ and electromagnetic observations providing information on the redshift $\left\{\hat{z} \right\}$ \cite{LIGOScientific:2017adf}. Consequently, we determine the joint posterior, $p(H_0 | \left\{x_{\rm GW}, \hat{z} \right\})$, for $N_{\rm obs}$ events in our population as follows
\begin{widetext}
\begin{equation}\label{Eq:jointposterior}
p(H_0 |\left\{x_{\rm GW}, \hat{z} \right\}) \propto \frac{p(H_0)}{N_{\rm s}(H_0)^{N_{\rm obs}}} \prod^{N_{\rm obs}}_{i=1} \int_{\Vec{\lambda}} \int_{D_L} p(x^{i}_{\rm GW}|D_L, H_0,\vec\lambda) p(\hat{z}_i|D_L, H_0) p(D_L) p(\Vec{\lambda}) dD_L d\vec{\lambda}, 
\end{equation}
\end{widetext}
\noindent
where $p(D_L)$ is the prior on the luminosity distance, and other waveform parameters are denoted by $\vec{\lambda}$, similar to the notation employed in Ref.~\cite{LIGOScientific:2017adf}. Hence, $p(\vec{\lambda})$ relates to priors for other GW parameters. Analogous to the approach used in Ref.~\cite{LIGOScientific:2017adf}, $N_s(H_0)$ incorporates the selection effect imposed by the finite sensitivity of GW detectors and is 
\begin{equation}
    N_s(H_0) = \int_{\rm det}  p(x_{\rm GW}|D_L, H_0,\vec\lambda) p(D_L) p(\vec{\lambda}) dD_L d\vec{\lambda} d{x_{\rm GW}}, 
\end{equation}
\noindent
integrated over the prior probability distributions of the parameters {$D_L, \vec{\lambda}$} and over ``detectable'' GW datasets. This factor is constructed on the assumption that the choice of whether or not an event is included in the analysis is a property purely of the observed data, $x_{\rm GW}$. In this analysis (as in~\cite{LIGOScientific:2017adf}) we assume that selection effects are dominated by the GW data; i.e., there are no selection effects for the EM counterpart. Additionally, we assume that GW detection is based on the observed SNR of the source.\footnote{In practice, we impose selection on the true, rather than the observed, luminosity distance of the source. This is the approach used widely in the literature. Selection on source parameters rather than observed data should really be treated differently~\cite{Gair:2022zsa}, but in practice this approximation does not change the results.} At the distances to which BNS events can be observed with current GW detectors, such a selection becomes a threshold in the luminosity distance of the source, which is what the GW detectors measure. As a consequence, the selection function has no dependence on the unknown cosmological parameters and so can be set to unity. 

Since our dataset lacks direct redshift measurements for each BNS merger event, we need to model this information. We consider all EM observations as securely detected and, hence, neglect a correction for EM selection effects. In the absence of actual electromagnetic observational data for redshifts, we simulate a redshift measurement $\hat{z}$ for the $i$th BNS merger event using a Gaussian probability density function

\begin{equation}\label{Eq:modelzobs}
    p(\hat{z}_i|D_L, H_0) = \frac{1}{\sqrt{2 \pi} \sigma_z} \exp \Bigl[ -\frac{1}{2} \Bigl( \frac{\hat{z} - z}{\sigma_z} \Bigr)^2 \Bigr]. 
\end{equation}
This allows us to associate each BNS source with a simulated measured redshift sampled from a Gaussian distribution centered on its true redshift value. We use the same Gaussian likelihood in Eq.~(\ref{Eq:jointposterior}) to obtain our Hubble constant posteriors. We examine the dependence of our analysis on the choice of the fractional error term, $C$, which contributes to the overall redshift uncertainty, $\sigma_z$, defined by the linear relationship $\sigma_z = z \cdot C$. This aims to quantify the sensitivity of our results to variations in the assumed fractional error value and, hence, in the precision of the EM measurement of the redshift. To infer joint constraints on $H_0$, we convert the measured redshift values to luminosity distances using Eq.~(\ref{Eq:linH0relation}). In our study, we do not explore different redshift uncertainty models as demonstrated in \cite{Turski:2023lxq}. Their work showed that the impact of different uncertainty models is much smaller than the current statistical errors. Hence, we expect that different redshift uncertainty models will not alter our results, which we will present in the next section.

\section{Results}\label{sec:results}

In accordance with the methodology detailed in Sec.~\ref{subsec:Bayframework}, we obtained combined $H_0$ estimates for each GW model. As described, we aim to investigate the sensitivity of our results to variations in the fractional redshift error value $C$, which we explore for $C=0.5\%$ and $3\%$. Our choice of the fractional error is driven by the inherent differences in redshift measurement techniques. Spectroscopic redshifts offer increased precision compared to photometric redshifts. Typical uncertainties for spectroscopic redshifts can be as low 0.01\% \cite{Steinhardt:2020kul}, while photometric redshift uncertainties can range up to 10\% or more \cite{Turski:2023lxq}. Hence, we intend to mimic a more precise spectroscopic and a less precise photometric measurement of the redshift. 

\begin{figure}
    \centering
    \includegraphics[width=\columnwidth]{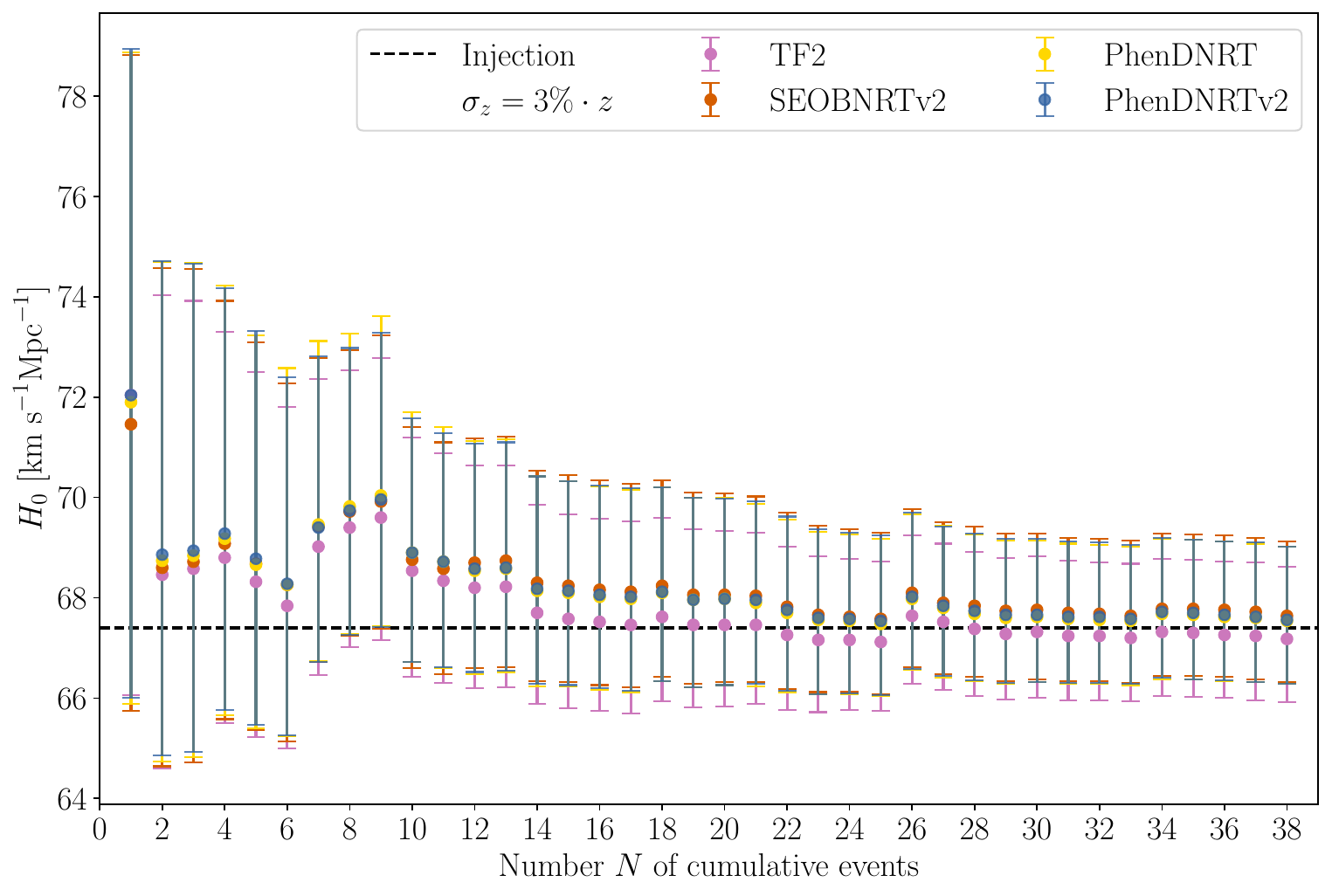}
    \caption{$H_0$ results versus the number of successively combined GW events for the GW models PhenDNRT (yellow), TF2 (purple), SEOBNRTv2 (orange), and the reference model PhenDNRTv2 (blue). Constraints on $H_0$ are reported at 95\% credible interval. The reference value of the Planck measurement \cite{Planck:2018vyg} is shown as a black dashed line. For this result, we modeled a $3$\% uncertainty on the true redshift as described in Sec.~\ref{subsec:Bayframework}.}
    \label{Fig:H0_zunc0_03_CI95}
\end{figure}

First of all, we study how our estimate on $H_0$ changes with the number of successively combined GW signals for each BNS event for $C = 0.03$, i.e., a 3\% fractional error in the modeling of the observed redshift. The result is shown in Fig.~\ref{Fig:H0_zunc0_03_CI95} and shows that all models recover the Planck measurement at $H_0 = 67.4~\rm kms^{-1}Mpc^{-1}$ within their uncertainty bands at 95\% credible interval. Consistent with theoretical expectations, incorporating more GW events gradually decreases the uncertainties associated with the $H_0$ estimates. Among 38 combined GW events, PhenDNRTv2 (the ``true'' model) and PhenDNRT align best with the Planck value, while SEOBNRTv2 and TF2 show slightly larger deviations. From Fig.~\ref{Fig:H0_zunc0_03_CI95}, we find that systematic biases in the extraction of the Hubble constant are below statistical uncertainties. 

\begin{figure}
    \centering
    \includegraphics[width=\columnwidth]{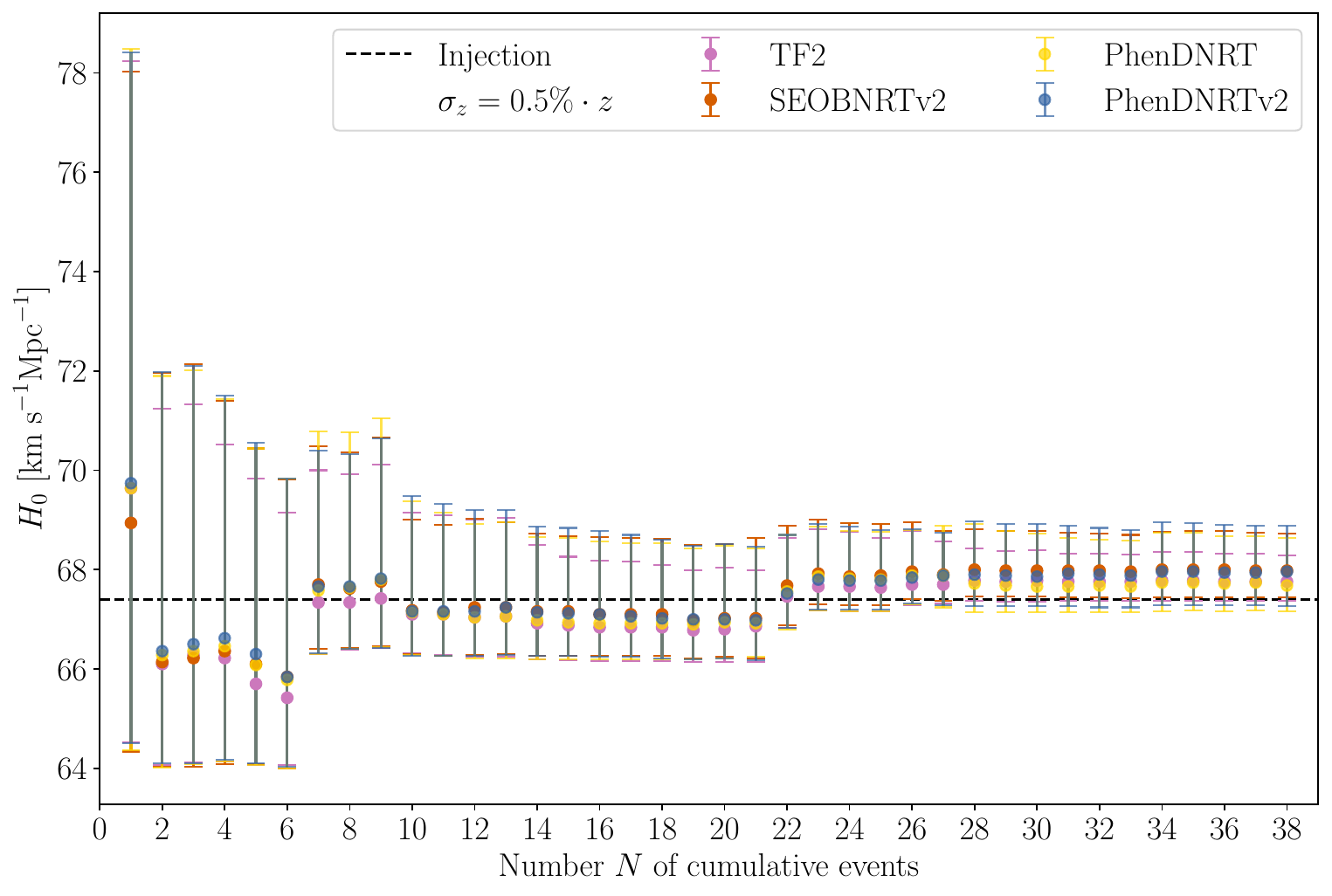}
    \caption{$H_0$ results versus the number of successively combined GW events for the GW models PhenDNRT (yellow), TF2 (purple), SEOBNRTv2 (orange), and the reference model PhenDNRTv2 (blue). Constraints on $H_0$ are reported at 95\% credible interval. The Planck measurement \cite{Planck:2018vyg} is shown as a black dashed line. For this result, we modeled a $0.5$\% uncertainty on the true redshift as described in Sec.~\ref{subsec:Bayframework}.}
    \label{Fig:H0_zunc0_005_CI95}
\end{figure}

Second, we show the results for a 0.5\% fractional error in Fig.~\ref{Fig:H0_zunc0_005_CI95}, i.e., assuming more precise EM measurements of the observed redshift. We find that our reference model, PhenDNRTv2, recovers the Planck measurement within 95\% credible interval. Likewise, analyses employing the GW models, PhenDNRT and TF2, recover this value when considering the combined data from all BNS events. The only exception is the model SEOBNRTv2 which does not recover the reference value when data from roughly 30 BNS events have been combined. Moreover, we find that there is a slight overestimation for all models when comparing with the Planck measurement. This is most likely due to a combination of statistical fluctuations and GW model-dependent differences in estimating the luminosity distance (as seen for TF2 in Fig.~\ref{Fig:H0_zunc0_03_CI95}). The initial selection of GW events within our sample introduces another source of statistical fluctuation. However, the present bias is not significant for our conclusions as the true value lies within the posterior uncertainty. 

At this point, it is crucial to acknowledge that our recovered $H_0$ values presented in Fig.~\ref{Fig:H0_zunc0_005_CI95} depend upon the specific random realization of observed redshifts generated through Eq.~(\ref{Eq:modelzobs}). This implies that any variation in the random realization of redshift measurements for each event would lead to a corresponding change in the inferred $H_0$ posterior. While we present a single realization in Fig.~\ref{Fig:H0_zunc0_005_CI95}, we investigated the dependence of our results concerning different random realizations of the redshift measurement. Notably, while maintaining a fixed fractional error of 0.5\%, we observed that the final, combined $H_0$ estimates exhibit a tendency to converge closer to the reference value of the Planck measurement under certain realizations (not shown in Fig.~\ref{Fig:H0_zunc0_005_CI95}).  

\begin{figure*}
    \centering
    \includegraphics[width=\textwidth]{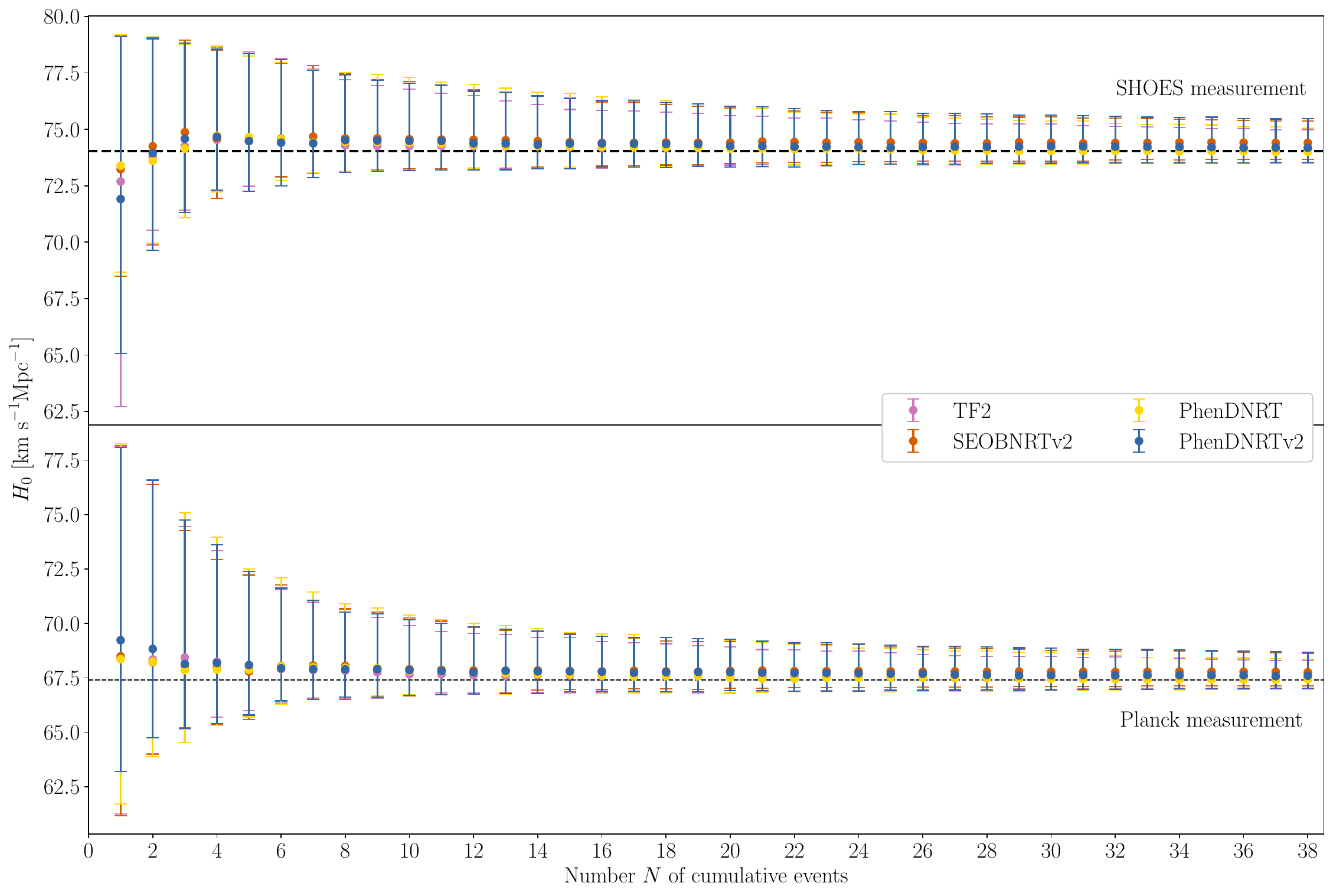}
    \caption{$H_0$ results versus the number of successively combined GW events for the GW models PhenDNRT (yellow), TF2 (purple), SEOBNRTv2 (orange), and the reference model PhenDNRTv2 (blue) for both, using the SHOES measurement of $H_0 = 74.03 \pm 1.42 \rm \ km \ s^{-1}Mpc^{-1}$ (top panel) ~\cite{Riess:2019cxk} and the Planck measurement of $H_0 = 67.4 \pm 0.5 \rm \ km \ s^{-1}Mpc^{-1}$ (bottom panel) ~\cite{Planck:2018vyg} as reference values. We use the same credible interval and fractional error on the redshift uncertainty as in Fig.~\ref{Fig:H0_zunc0_005_CI95}, but randomly shuffled the order of BNS events in our sample for 100 times, while employing 100 random realizations of redshift measurements.}
    \label{Fig:H0_zunc0_005_CI95_Nshuffle_100}
\end{figure*}

In order to minimize the impact of stochastic redshift realizations and of specific BNS event orderings in our sample, we implement a double randomization procedure.\footnote{Despite this procedure, our GW posteriors are not randomized, which can lead to slight offsets or biases as we can see in our study.} In this procedure, we draw $100$ random redshift values assuming a Gaussian distribution with a $0.5\%$ uncertainty for each BNS event within the sample. Equation (\ref{Eq:jointposterior}) is then used to compute the combined $H_0$ value for $100$ times. In each iteration, the order of the BNS events is randomized, and a different drawn redshift value is used for each event. This procedure yields $100$ independent realizations of the combined $H_0$ posterior distribution. Finally, we compute the median of the distribution to extract the final results of $H_0$. The results are shown in Fig.~\ref{Fig:H0_zunc0_005_CI95_Nshuffle_100}, for which we modeled a $0.5\%$ uncertainty on the true redshift. Our result depicts a convergence towards the Planck value for all GW models. Consistent with prior findings, it also reveals a progressive uncertainty reduction and that all GW models recover the Planck measurement. 

From Figs.~\ref{Fig:H0_zunc0_005_CI95} and \ref{Fig:H0_zunc0_005_CI95_Nshuffle_100}, we find that PhenDNRT tends to yield smaller $H_0$ values as compared to PhenDNRTv2. This could result from the fact that the PhenDNRTv2 includes an amplitude tidal correction term \cite{Dietrich:2019kaq}, while this is not the case for PhenDNRT. Because of this correction, the amplitude for PhenDNRT waveforms will be larger as compared to PhenDNRTv2. In order to compensate, PhenDNRT needs to go to higher luminosity distances and, hence, will predict smaller values for the Hubble constant. Another distinctive feature is that the uncertainty is smallest for TF2 when all data has been taken into account. To investigate this observation, we computed the Fisher information matrix (FIM) for both TF2 and PhenDNRTv2 across all events using \textsc{GWFAST}~\cite{Iacovelli:2022bbs, Iacovelli:2022mbg}. The FIM analysis revealed that 55\% of the events exhibited a lower distance error for TF2 as compared to PhenDNRTv2. This finding suggests that TF2 can achieve a narrower posterior width or lower variance. However, we would still expect it to have a larger bias, which would eventually dominate the uncertainty once sufficiently many events have been combined. 
Moreover, we repeat the analysis employing the Hubble constant measurement reported by the SHOES Collaboration~\cite{Riess:2019cxk} as the reference value, $H_0 = 74.03 \pm 1.42 \rm \ km \ s^{-1}Mpc^{-1}$. Figure \ref{Fig:H0_zunc0_005_CI95_Nshuffle_100} shows that we obtain consistent results compared to the scenario where we utilize the Planck measurement. Consequently, our primary conclusions remain unchanged.

Overall, our results indicate that waveform model-dependent biases on the extraction of the Hubble constant are small for current detectors. Ascribing the subtle discrepancies observed in the combined $H_0$ estimates solely to specific model descriptions or assumptions presents a significant challenge. First, GW models only possess a limited capability to accurately infer the distance and, consequently, the Hubble constant, so we are dominated by statistical rather than systematic uncertainties. Second, our employed GW models exhibit interdependencies within specific regimes, which hamper an unambiguous explanation attributable to individual models.

Finally, we use our results compared to the Planck measurement shown in Fig.~\ref{Fig:H0_zunc0_005_CI95_Nshuffle_100} to estimate how many cumulative BNS events will be required to achieve a statistical uncertainty that falls below the waveform model systematics. To understand how much the final measurement of the Hubble constant might be affected by the choice of a certain model, we calculate the difference between the final $H_0$ value obtained using our reference model PhenDNRTv2 and the final values obtained using all the other models. The largest systematic shift of $\Delta H_0 = 0.18 \rm \ km \ s^{-1}Mpc^{-1}$ is present when comparing to PhenDNRT, while the final result of TF2 has the smallest shift with $\Delta H_0 = 0.08 \rm \ km \ s^{-1}Mpc^{-1}$. As the statistical uncertainty drops as $\sim 1/\sqrt{N_{\rm obs}}$, we can estimate the number of BNS events required to fall below the above-reported values. For our reference model PhenDNRTv2 to possess statistical uncertainties smaller than the systematic shift of $\Delta H_0 = 0.18 (0.08) \rm \ km \ s^{-1}Mpc^{-1}$ roughly 3500 (16,500) BNS observations would be required with the current generation of GW detectors.

\section{Conclusions}\label{sec:conclusion}

In this study, we investigated the impact of gravitational waveform systematics on the extraction of the Hubble constant. Employing the bright siren approach, we obtained combined estimates on the Hubble constant by using previous simulated GW data from a population containing 38 BNS sources published in \cite{Kunert:2021hgm} and by modeling a corresponding EM observation to provide information on the redshift. Overall, our analysis reveals that the uncertainty in the luminosity distance arising from GW models exerts a negligible influence on the determination of the Hubble constant for the current generation of GW detectors. However, our analysis is limited because our employed GW models are based on the quadrupolar mode and do not include any higher-order modes.
While current systematic biases on the Hubble constant are below statistical uncertainties, this picture will change with upcoming third generation detectors like the Einstein Telescope (ET) \cite{Punturo_2010, Maggiore:2019uih, Branchesi:2023mws} or the Cosmic Explorer (CE) \cite{Reitze:2019iox, evans2021horizon}. Projected BNS detection rates of $10^4$ BNS/year for ET \cite{Regimbau:2012ir, Singh:2021zah, Branchesi:2023mws} or $~10^5$ BNS/year for a network consisting of ET and CE \cite{2024arXiv240102604W} imply diminishing statistical uncertainties, necessitating an increased focus on reducing systematic biases in the future.

\acknowledgements

N.K., P.T.H.P., and T.D.\ acknowledge support from the Daimler and Benz Foundation for the project ``NUMANJI''. T.D.\ acknowledges support from the European Union (ERC, SMArt, 101076369). Views and opinions expressed are those of the authors only and do not necessarily reflect those of the European Union or the European Research Council. Neither the European Union nor the granting authority can be held responsible for them. P.T.H.P. is supported by the research program of the Netherlands Organisation for Scientific Research (NWO). 

We acknowledge usage of computer time on Lise/Emmy of the North German Supercomputing Alliance (HLRN) (Project No. bbp00049), on HAWK at the High-Performance Computing Center Stuttgart (HLRS) (Project No.~GWanalysis 44189), and on SuperMUC NG of the Leibniz Supercomputing Centre (LRZ) (Project No. pn29ba).

\bibliography{main.bib}

\end{document}